\documentclass[conference]{IEEEtran}
\ifCLASSINFOpdf
\else
\fi
\usepackage{amsfonts}
\usepackage{amsmath,amsthm,txfonts}
\usepackage{graphicx,times,amsmath,color, varioref, subfigure} % Add all your packages here

\IEEEoverridecommandlockouts

\begin{document}
\title{Comparison of Distance Metrics for Hierarchical Data in Medical
Databases \thanks{Diman Hassan, Uwe Aickelin and Christian Wagner are with the School of Computer Science, University of Nottingham, Nottingham, United Kingdom (email: \{dsh, uxa, cxw\}@cs.nott.ac.uk).}}

\author{Diman Hassan, Uwe Aickelin and Christian Wagner}


%Telephone: (800) 555--1212, Fax: (888) 555--1212}


\maketitle

\begin{abstract}
Distance metrics are broadly used in different research areas and applications, such as bio-informatics, data mining and many other fields. However, there are some metrics, like $pq$-gram and Edit Distance used specifically for data with a hierarchical structure. Other metrics used for non-hierarchical data are the geometric and Hamming metrics. We have applied these metrics to The Health Improvement Network (THIN) database which has some hierarchical data. The THIN data has to be converted into a tree-like structure for the first group of metrics. For the second group of metrics, the data are converted into a frequency table or matrix, then for all metrics, all distances are found and normalised. Based on this particular data set, our research question: which of these metrics is useful for THIN data?. This paper compares the metrics, particularly the $pq$-gram metric on finding the similarities of patients' data. It also investigates the similar patients who have the same close distances as well as the metrics suitability for clustering the whole patient population. Our results show that the two groups of metrics perform differently as they represent different structures of the data. Nevertheless, all the metrics could represent some similar data of patients as well as discriminate sufficiently well in clustering the patient population using $k$-means clustering algorithm.
\end{abstract}

% no key words

\section{Introduction}
\label{sec:section1}

\PARstart{S} {ince} the representation of structured objects in large and modern databases like The Health Improvement Network (THIN) database becomes more complex and important, such structures should be considered when searching for similar objects. Therefore, finding an efficient measurement for discovering similar objects in data sets is the key feature when the task is to classify new objects or to cluster data objects. The $pq$-gram \cite{Augsten2010} and Edit Distance \cite{Kailing2004} metrics are known to be two good approaches that have been used to measure the similarity of the structured data objects, especially in Trees. The limitation of Edit Distance metric is related to the computational complexity which is considered very high \cite{Zhang1989} as compared to the $pq$-gram distance metric. \\
On the other hand, there are other metrics that are simple and implemented on non-structured data, such as Euclidean, Minkowski, Manhattan and Hamming Distance metrics \cite{cordeiro2012}. Some of these metrics have been compared to other measures to find their efficiency. In \cite{shahid2009}, a comparison has been made between the geometric metrics and actual measures to estimate the distance in spatial analytical models. The results gave accurate distances for the actual distances than to the geometric metrics. 
Recently, using THIN database (www.thin-uk.com) which is belong to the general practice electronic healthcare database, some research \cite{reps2012} \cite{reps2011} have been performed using data mining techniques, such as association and sequential patterns. The purpose was to detect association between patient attributes (e.g. age, gender, medical history) and adverse events of drugs. No other data mining technique has been applied to the THIN database yet, such as clustering; this motivated us to use the unexplored clustering approach for the prediction and detection of negative side effects of drugs. The overarching aim of our research is to cluster hierarchical data to identify adverse side effects of drugs in the THIN database. However, clustering techniques need distance measures to represent the similarity between patients who have similar side effects. For this reason, this preliminary work aims to find the useful and suitable measure for our hierarchical data set in order to cluster patients. To achieve this aim, different metrics are considered and applied to the THIN data and their results compared. The investigation determines if these metrics can measure similarity and find similar patients (i.e. the patients who have similar side effects of drugs). Additionally, by looking at the whole patient population, is any of the metrics able to accurately represent similarity between patients when using, for example the $k$-means clustering algorithms \cite{jain1988}?.\\
The layout of this paper is as follows. In Section~\ref{sec:section2}, a background on the THIN database and the distance metrics is given. The data preparation for both groups of metrics, the calculation of the distances and the clustering using those metrics are explained in Section~\ref{sec:section3} followed by a discussion on the results in Section~\ref{sec:section4}. Section~\ref{sec:section5} presents a summary and the conclusion of the work.

\section{MATERIALS AND METHODS}
\label{sec:section2}
\subsection{Background on THIN Database}
The THIN database is one of the electronic health-care longitudinal databases that contains anonymous electronic medical records extracted directly from general practices throughout the United Kingdom. The database contains information of each patient registered within the general practice including personal details, such as gender, date of birth, date of registration and family history. In addition, the data on all the drug prescriptions and the associated set of symptoms based on which the drug is prescribed are also included. The individual medical record is represented in the THIN database by a reference code named as read code. The latter is an alphanumeric code that defines and groups illnesses using the hierarchical nosology system. The read codes are also comprehensive coded medical language developed in the UK and funded by the National Health Service (NHS). In this paper, we test our experiments on a group of patients between the age of 0 and 17 years old. The information shown in Table \ref{table1} was extracted from THIN for two kinds of drugs that have been chosen based on the number of prescriptions. The first drug DESLORATADINE has a large number of prescriptions and is used to treat allergies under the group of Antihistamines. The second drug has a smaller number of prescriptions and belongs to the family of Tricyclics that relate to antidepressant drugs \cite{joint2012}. For our experiments, a sample size of 9949 prescriptions after 30 days of taking the drug (representing 988 patients) out of 53,995 prescriptions (representing 18,293 patients) have been tested to find the similarity between them for the first drug.  For the second drug we used all the prescriptions (1172) after 30 days of taking the drug for 42 patients.
\begin{table}[!h]
\centering
\caption{A subset of information from the database for two kinds of drugs}
\label{table1}
\begin{tabular}[width=1cm]{|c|c|c|}
\hline
\hline
        & DESLORATADINE &  DOXEPIN\\
\hline
\hline
All drug's codes in THIN data set & 6 & 15\\
\hline
All prescription & 358,768 & 72448\\
\hline
All patients & 81,000 & 6152\\
\hline
All prescription(0-17 years) & 53,995 & 2014\\
\hline
All patients (0-17 years) & 18,293 & 60\\
\hline
All presc.(0-17) after 30 days & 9949 & 1172\\
\hline
All Patients(0-17) after 30 days & 988 &42\\
\hline
\end{tabular}
\end{table}
\subsection{Background on Distance Metrics}
A metric space ($X$, $d$) is a set $X$ that has the concept of distance $d$($x$, $y$) between any pair of points $x$, $y$ $\in X$ and the metric is a function on the set $X$ that satisfies the following properties for a distance \cite{oxtoby1971} \cite{korner2010}.\\ \emph{Definition}: a metric $d$ on a set $X$ is a function $d$: $X \times X \rightarrow \mathbb R$ such that for all $x$, $y$ $\in X$:\\
$d($x$, $ y$) \geq 0  \forall  $ x$, $ y $ \in X$. (Non-negativity).\\
$d($x$, $ y$) $ = $ 0 \iff $x$ $ = $ $y$ $ (Identity).\\
$d($x$, $ y$) $ = $ d($y$, $ x$) \forall $ x$, $ y $ \in X$. (Symmetry).\\
$d($x$, $ y$) \leq d($x$, $ z$) $ + $ d($z$, $ y$ )$. (Triangle inequality) $\forall $ x$, $ y $ and $ z $ \in X$.\\
The following are the six distance metrics used in this study:\\
\subsubsection{Euclidean Distance Metric}
 Euclidean metric is a distance $d$ on the space $\mathbb R^n \times \mathbb R^n \rightarrow \mathbb R$  which is defined as a distance between any two points in space  $\mathbb R^n \times \mathbb R^n \rightarrow \mathbb R$
%------------------------------------------------------
\begin{equation}
d(x,y) = \sqrt{(x_1 - y_1)^2 + (x_2 - y_2)^2 + ... + (x_n - y_n)^n}
\label{eq:equation1}
\end{equation} where $x$ = ($x_1$,  $x_2$, ... ,$x_n$)   , $y$ = ($y_1$, $y_2$, ... ,$y_n$) \cite{gower1982}.\\
%------------------------------------------------------
\subsubsection{Minkowski Distance Metric}
Minkowski metric is a $p$-metric between $n$-dimensional points $x$ $=$ ($x_i$) and $y$ $=$ ($y_i$) defined as:
%------------------------------------------------------
\begin{equation}
\label{eq:equation2}
d(x,y) = \sqrt[p]{\sum_{i=1}^n |(x_i - y_i)^p|}
\end{equation}
%------------------------------------------------------
If $p=2$, it is called Euclidean distance and if $p=1$ it is called Manhattan or city block distance. If $p=\infty$, then it is called Chebyshev or maximum distance \cite{cordeiro2012}. In our experiment, $p$ $=$ 3 has been used.\\

\subsubsection{Manhattan Distance Metric}
It is a special case of the Minkowski metric when $p = 1$ \cite{cordeiro2012}:
%----------------------------------------------------------------------------
\begin{equation}
d(x,y) = \sum_{i=1}^n |(x_i - y_i)| 
\label{eq:equation3}
\end{equation}
%----------------------------------------------------------------------------
where $x$ $=$ ($x_1$, $x_2$,..., $x_n$) and $y$ $=$ ($y_1$, $y_2$,..., $y_n$) \\

\subsubsection{Hamming Distance Metric}
Hamming distance is used for the detection and correction of errors in digital communications. It is defined as the number of different symbols between two equal length sequences. For example, the hamming distance between "\textcolor{red}{t}o\textcolor{red}{n}e\textcolor{red}{d}" and "\textcolor{red}{r}o\textcolor{red}{s}e\textcolor{red}{s}" is 3 and between 21\textcolor{red}{7}3\textcolor{red}{8}9 and 21\textcolor{red}{3}3\textcolor{red}{7}9 is 2 \cite{hosangadi2012}. \\

\subsubsection{Edit Distance Metric}
According to Kialing et al. \cite{Kailing2004}, the definition of the Edit Distance measure between two trees $T_1$ and $T_2$ is the minimum cost of all edit sequences that transform $T_1$ to $T_2$: Edit Distance($T_1$, $T_2$) = $\min\{c(S)\backslash{S}$ a sequence of edit operations transformations $T_1$ into $T_2\}$. Kialing et al. claimed the advantage of using the edit distance as a similarity measure provided the mapping between the nodes in two trees during the term of edit sequence (Insertion, Deletion and Relabeling nodes in a tree $T$).\\

\subsubsection{PQ-Gram Distance Metric}
\begin{figure}[!h]%
\includegraphics[width=7cm]{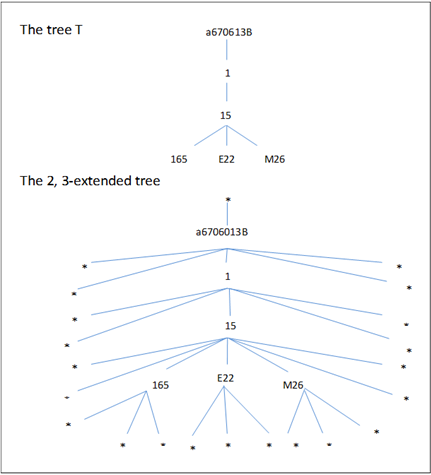}%
\centering
\caption{An example of a tree $T$ and its 2, 3-Extended tree}%
\label{figure1}%
\end{figure}

\begin{figure}[!h]
\centering
\includegraphics[width=8cm]{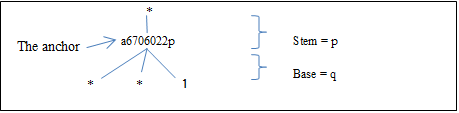}%
\caption{An example of single $pq$-gram from a THIN data tree}%
\label{figure2}%
\end{figure}

The $pq$-gram distance has been proposed by Augsten et al. \cite{Augsten2010} and is mainly used for computing distances between ordered labeled trees. The $pq$-grams of a tree are all its sub-trees of a specific shape. The specific shape of the $pq$-gram is based on the values of two parameters $p$ and $q$. The tree $T$ shown in Fig. \ref{figure1} is expanded by inserting dummy nodes (*) to make sure that each node appears at least in one $pq$-gram. The expansion of each tree is done by inserting $p$-1 before the root node, insert $q$-1 before the first and after the last child of each non-leaf node and insert $q$ nodes to each leaf node, for example $p$ = 2, $q$ = 3 in Fig. \ref{figure2}. After the expansion process, the 2, 3-grams are extracted to produce the list of $pq$-grams. An example of a single 2, 3-gram is given in Fig. \ref{figure2} where $p$ = (*, a6706022p) is the stem and $q$ = (*, *, 1) is the base. The trees that have a large number of common $pq$-grams are considered more similar than those trees that have less; Furthermore, the $pq$-gram distance is used to approximately match hierarchical data of large sources using the following equations:
%------------------------------------------------------
\begin{equation}
\emph{dist}^{(p,q)} (T_1, T_2) = |I_1 \uplus I_2| - 2|I_1 \nplus I_2|
\label{eq:equation4}
\end{equation} 
%------------------------------------------------------
Where $T_1$, $T_2$ are the two trees, and $p$ and $q$ are the two parameters that specify the shape of the $pq$-gram. The $pq$-gram indexes, $I_1$ and $I_2$ are the bags of  
Label-tuples of all $pq$-grams of $T_1$ and $T_2$, respectively. In addition, the $\uplus$ refers to the bag union between $I_1$ and $I_2$ and the $\nplus$ refers to the bag intersection between the same indexes.
The normalisation of the $pq$-gram distances is as follows:
%------------------------------------------------------
\begin{equation}
dist\_norm^{(p, q)} (T_1, T_2) = \frac{dist^{(p,q)} (T_1,T_2)}{|I_1 \uplus I_2|-|I_1 \nplus I_2|}       
\label{eq:equation5}
\end{equation} \\
%-------------------------------------
The $pq$-gram metric has been proposed originally to approximately match similar hierarchical information from autonomous sources that may have different representation in the sources \cite{Augsten2010}. The $pq$-gram metric has the advantage of computational efficiency and can be computed in $\mathcal{O}(n \log n)$ time and $\mathcal{O}(n)$ space. Another advantage of the $pq$-gram distance is that it can be tuned by adjusting the two parameters $p$ and $q$ \cite{srivastava2010}. The determination of $p$ and $q$ values depends on the underlying semantics of the data. In general, increasing the values of $p$ and $q$ makes the distance between two trees more sensitive to the structure of the trees rather than to the data, while decreasing them makes the distance sensitive to the data. As an example, in our experiments we have used different values of $p$ and $q$: for $p$ = 1 and $q$ = 3 and for $p$ = 2, $q$ = 3, the results of $pq$-gram distances are shown in Table \ref{table3} and Table \ref{table5}. The results reveal that better distances are obtained when $p$ = 1 and $q$ = 3. 
%===========================================================
%====================================================
%====================================================

\section{EXPERIMENTS AND RESULTS}
\label{sec:section3}
\subsection{Data Preprocessing}
\begin{figure}[!h]%
\centering
\includegraphics[width=7cm]{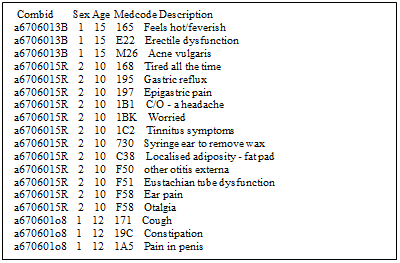}%
\caption{Part of the THIN data extracted based on specific attributes}%
\label{figure3}%
\end{figure}

The THIN data is converted into trees before applying the $pq$-gram and Edit Distance metrics, while the data for the geometric and Hamming distance metrics is converted into a frequency table. The data extracted from THIN is based on different patient's attributes such as the patient's unique ID, the gender, the age of first taking the specified drug and the medical codes related to the drug. The medical events are chosen at level 3 (the first three digits of the read codes like H33). Fig. \ref{figure3} shows part of this information represented in THIN for three patients which have unique identifiers in the database (a6706013B, a6706015R, a670601o8):\\
\subsubsection{PQ-Gram and Edit Distance Preparation}
\begin{figure}[!h]%
\centering
\includegraphics[width=8cm]{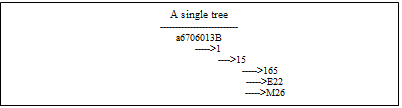}%
\caption{A tree representation from THIN data}%
\label{figure4}
\end{figure}

\begin{figure}[htp]%
\centering
\includegraphics[width=7cm]{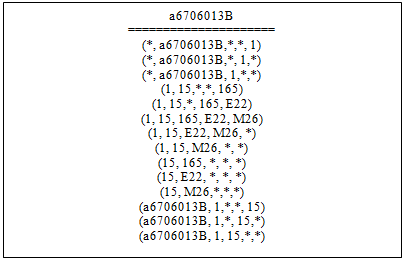}%
\caption{The 2, 3-grams of a tree $T$}%
\label{figure5}%
\end{figure}

From the data in Fig. \ref{figure3}, we have converted each patient's records into a tree as depicted in Fig. \ref{figure4} to enable the computation of both $pq$-gram and Edit Distance metrics. For the $pq$-gram metric each tree is expanded in the same way as in Fig. \ref{figure1}. For our experiments, we use ($p = 1$, $q = 3$) and ($p = 2$, $q = 3$). After the process of tree expansion, the $pq$-grams are extracted for each tree; Fig. \ref{figure5} shows the 2, 3-grams for the tree in Fig. \ref{figure4}. The $pq$-gram distance between two trees is formed by all the common $pq$-grams between them and computed using equation \eqref{eq:equation4}, while the calculation of the distances for the Edit Distance is performed by inserting, deleting or re-labeling nodes to convert one tree to another. The single edit operation has cost 1 and the Edit Distance between two trees is equal to the minimum cost or minimum number of edit operations to convert one tree to another.\\

\subsubsection{Geometric and Hamming Metrics preparation}
The THIN data for the Euclidean, Minkowski, Manhattan and Hamming metrics has been converted into a frequency table as shown in Table \ref{table2}. The table represents how many times each patient had a specific symptom after taking the specified drug.
%=====================================================
\begin{table*}[t]
\centering
\caption{The frequency table from THIN data}
\label{table2}
\centering
\setlength{\tabcolsep}{5pt}
\begin{tabular}{|c|c|c|c|c|c|c|c|c|c|c|c|c|c|c|c|c|c|c|c|c|}
\hline
 Patient's ID & \multicolumn{15}{|c|}{The medical events} & \multicolumn{4}{|c|}{Patient's ages}&Sex\\ 
\hline
   &  168 & 171 & 195  & 19C & 1A5 & 730 &F58& H17 & M0.& M26 & N24 & N32 & SD. & SL. & ZL5 & 10 & 11 & 12 & 15& \\
\hline
 a6706013B &  0 & 0 & 0 & 0 & 0 & 0 & 0 & 0 & 0 & 1 & 0 & 0 & 0 & 0 & 0 & 0  & 0  &  0 & 1 & 1 \\      

 a6706015R & 1 & 0 & 1 & 0 & 0 & 1 & 2 & 0 & 0 & 0 & 0 & 0 & 0 & 0 & 0 & 1  & 0  &  0 & 0 & 2 \\      

a670601o8 &  0 & 1 & 0  & 1 & 1 & 0 & 0 & 1 & 1 & 1 & 1 & 1 & 1 & 1 & 0 & 0  & 0  &  1 & 0 & 1 \\      

 a670601yJ &  0 & 0 & 0 &  0 & 0 & 0 & 0 & 0 & 0 & 1 & 0 & 0 & 0 & 0 & 1 & 0  & 1  & 0 & 0 & 2 \\      
\hline
\end{tabular}
\end{table*}
%===================================================
The table also contains additional columns, one for the patient's gender (In THIN, 1 = male, 2 = female) and others for the different ages of each patient taking the drug. In Table \ref{table2}, the ages of the patients are 10, 11, 12 and 15.\\

\subsubsection{Distances Calculation}
The distances using all the six metrics applied to the THIN data are calculated and normalised. The normalisation of the distances is to demonstrate that the small distances that are close to 0 indicate similar patients, while the large distances that are near to 1 indicate dissimilar patients. In the case of Euclidean, Minkowski and Manhattan metrics, the data in Table \ref{table2} has been used to calculate the distances using equations \eqref{eq:equation1}, \eqref{eq:equation2} and \eqref{eq:equation3}, respectively. For the calculation of Hamming distances, the number of different values between two of equal length sequences from Table \ref{table2} has been taken into account. The normalisation of the distances has been calculated using the formula: \emph{norm-dist.} \emph{$(x) = x - \min(x) / \max(x) - \min(x)$} where $x$ refers to the distance between two patients. Regarding the $pq$-gram metric, the distance between two trees of patients is defined as a symmetric difference between the two sets of $pq$-grams using equation \eqref{eq:equation4}, while the normalisation of the $pq$-gram distances is calculated using equation \eqref{eq:equation5}. On the other hand, the Edit Distance distances are equal to the minimum number of edit operations (insert, delete or rename nodes) when converting one tree to another. Each edit operation has cost 1 and based on the distance being equal to the minimum cost of converting $T_1$ to $T_2$. The Tree Edit Distance Normalisation (TED\_NORM) is:\\
\begin{equation}
TED\_NORM (T_1, T_2) = \frac{TED(T_1, T_2)}{(|T_1| + |T_2|)} \
\label{eq:equation6}              
\end{equation}\\
Where $(|T1|+|T2|)$ means the sum of the two trees' nodes. The results of calculating the distances using all the six metrics are summarised in Table \ref{table3} and Table \ref{table5} for DESLORATADINE and DOXEPIN, respectively. The tables contain all the smallest normalized distances for patients (the most similar data) among the other distances.\\ 
The results for the first drug show that geometric and hamming metrics could find similar patients as the distance between two patients equal to zero. In contrast, the $pq$-gram and Edit Distance metrics produced a very few similar patients, like (a670605Up, a670602uS) and (a67340327, a681001KN) besides others who have some similarity or close distances to the identical level between patients. The reason behind that is related to the structure of the data which is a hierarchical tree structure. \\
On the other hand, the experiment for the second drug also produced a number of similar patients in their medical events based on the geometric and hamming metrics as shown in Table \ref{table5}, while for the $pq$-gram and Edit Distance metrics the table shows no similar distances. The reason behind that could be the lack of data for the second drug. 

\begin{table*}[t]
\renewcommand{\arraystretch}{1}
\centering
\caption{Smallest normalised distances for patients taking DESLORATADINE drug}
\label{table3}
\centering
\begin{tabular}{|c|c|c|c|c|c|c|c|}
\hline
\multicolumn{8}{|c|}{The Normalised Distances}\\
\hline
 patient's ID & Euclidean & Minkowski& Manhattan & Hamming & Edit Distance & 1, 3-Grams & 2, 3-Grams \\
\hline
 a670605Up, a670602uS & 0	& 0	& 0	& 0	& 0.25 & 0 & 0\\
\hline
 a6732002X, a673200WF	& 0	& 0	& 0	& 0	& 0.25 & 0.888889 & 1\\
\hline
a6732002X, a673201@y	& 0	& 0	& 0	& 0	& 0.25 & 0.888889 & 1\\
\hline
a673200tm, a673201j7	& 0	& 0	& 0	& 0	& 0.25 & 0.888889 & 1\\
\hline
a673201@y, 673200WF	& 0	& 0	& 0	& 0	& 0.25 & 0.888889 & 1\\
\hline
a673201Wt,a6732025y	& 0	& 0	& 0	& 0	& 0.25 & 0.888889 & 1\\
\hline
a673201wI, a678701pI	& 0	& 0	& 0	& 0	& 0.6666 & 0.888889 & 1\\
\hline
a67340327, a681001KN	& 0	& 0	& 0	& 0	& 0.25 & 0 & 0 \\
\hline
a677505bO, a677505pe	& 0	& 0	& 0	& 0	& 0.25 & 0.888889 & 1\\
\hline
a683104@Y, 677505bO	& 0	& 0	& 0	& 0	& 0.6666 & 0.888889 & 1\\
\hline
a683104@Y, a677505pe	& 0	& 0	& 0	& 0	& 0.6666 & 0.888889 & 1\\
\hline
a673201wI, a777805mH	& 0	& 0	& 0	& 0	& 0.25 & 0.888889 & 1\\
\hline
a673402zw, a683105Bk	& 0	& 0	& 0	& 0	& 0.25 & 0.888889 & 1\\
\hline
a678701pI, a777805mH	& 0	& 0	& 0	& 0	& 0.25 & 0.888889 & 1\\
\hline
a791600uB,a777806FG	& 0	& 0	& 0	& 0	& 0.25 & 0.888889 & 1\\
\hline
a7916065T, a777800Gj	& 0	& 0	& 0	& 0	& 0.25 & 0.888889 & 1\\
\hline
\end{tabular}
\end{table*}

\begin{table*}[t]
\renewcommand{\arraystretch}{1}
\centering
\caption{The Shared Medical Events for Patients in Table \ref{table3}}.
\label{table4}
\centering
\begin{tabular}{|c|c|c|c|}
\hline
 Patient's ID & The medical events For the first patient & The medical events for the second patient & The description of the event  \\
\hline
 a670605Up, a670602uS & 1B8	& 1B8	& Itchy eye symptom	\\
\hline
 a6732002X, a673200WF	& 17Z	& 17Z	& Respiratory symptom NOS	\\
\hline
a6732002X, a673201@y	& 17Z	& 17Z	& Respiratory symptom NOS	\\
\hline
a673200tm, a673201j7	& ZL5	& ZL5	& Referral to orthopaedic surgeon	\\
\hline
a673201@y, 673200WF	  & 17Z	& 17Z	& Respiratory symptom NOS	\\
\hline
a673201Wt,a6732025y	  & 740	& 740	& Submucous diathermy to turbinate of nose	\\
\hline
a673201wI, a678701pI	& H05	& H05	& Upper respiratory tract infection NOS	\\
\hline
a67340327, a681001KN	& 171	& 171	& Cough	\\
\hline
a677505bO, a677505pe	& 8B3	& 8B3	& Medication requested	\\
\hline
a683104@Y, 677505bO	  & 8B3	& 8B3	& Medication requested	\\
\hline
a683104@Y, a677505pe	& 8B3	& 8B3	& Medication requested	\\
\hline
a673201wI, a777805mH	& H05	& H05	& Upper respiratory tract infection NOS \\
\hline
a673402zw, a683105Bk	& H17, 8B3	& H17, 8B3 & Hay fever or pollens, Medication requested	\\
\hline
a678701pI, a777805mH	& H05	& H05	& Upper respiratory tract infection NOS	\\
\hline
a791600uB,a777806FG  	& H17	& H17	& Hay fever or pollens	\\
\hline
a7916065T, a777800Gj	& A78 & A78	& Verrucae warts or Molluscum contagiosum	\\
\hline
\end{tabular}
\end{table*}

\begin{table*}[t]
\renewcommand{\arraystretch}{1}
\centering
\caption{Smallest normalised distances for patients taking DOXEPIN drug}
\label{table5}
\centering
\begin{tabular}{|c|c|c|c|c|c|c|c|}
\hline
\multicolumn{8}{|c|}{The Normalised Distances}\\
\hline
 patient's ID & Euclidean & Minkowski& Manhattan & Hamming & Edit Distance & 1, 3-Grams & 2, 3-Grams \\
\hline
 a793901c8,a9910027z & 0 & 0	& 0 & 0 & 0.4761 & 0.971 & 0.967\\
\hline
b977401S1,a999104cU	& 0 & 0	& 0 & 0 & 0.4 & 0.923 & 1	\\
\hline
g989501KB,a999104cU	& 0 & 0	& 0 & 0 & 0.375 & 1 & 1 \\
\hline
g989501KB,b990804AL	& 0 & 0	& 0 & 0 & 0.5 & 1 & 1	\\
\hline
b990804AL, a999104cU & 0 & 0	& 0 & 0 & 0.375  & 1 & 1\\
\hline
\end{tabular}
\end{table*}

\begin{figure*}
  \centering
  
    \subfigure[The clusters of patients using Euclidean metric, DECLORATEDINE drug \label{figure6a}]{\includegraphics[width=3in]{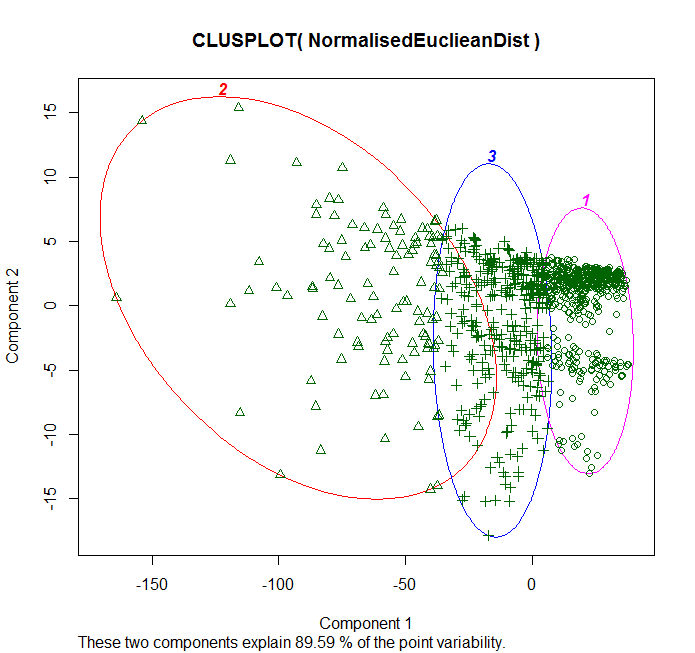}}\qquad
    \subfigure[The clusters of patients using $pq$-gram metric, DECLORATEDINE drug \label{figure7b}]{\includegraphics[width=3in]{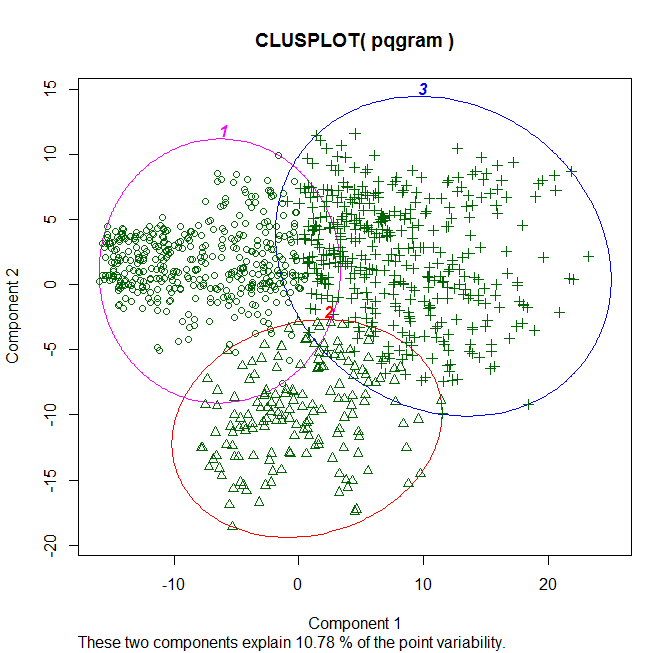}}\qquad
    \subfigure[The clusters of patients using Euclidean metric, DOXEPIN drug \label{figure8c}]{\includegraphics[width=3in]{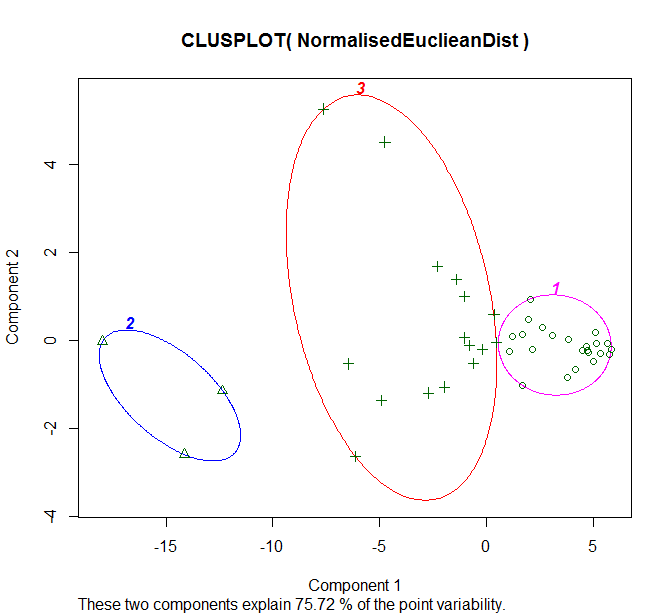}}\qquad
    \subfigure[The clusters of patients using $pq$-gram metric, DOXEPIN drug \label{figure9d}]{\includegraphics[width=3in]{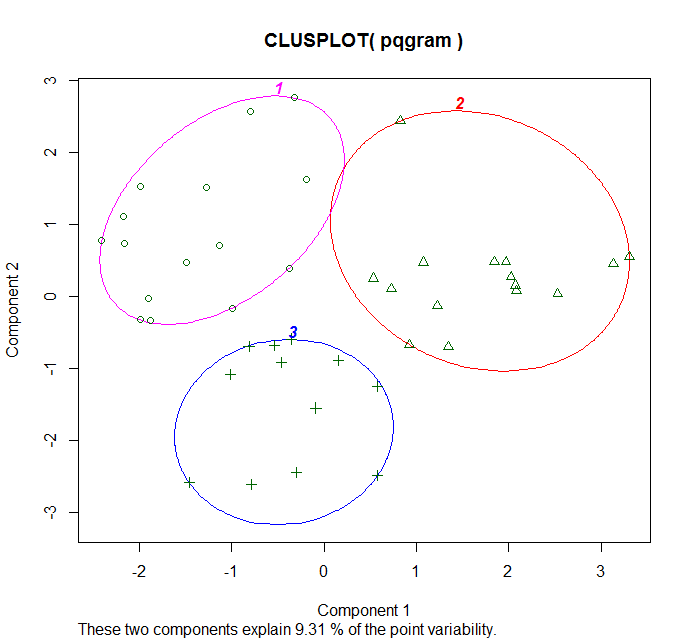}}

  \caption{The clusters of patients using Euclidean and $pq$-gram metrics}
  \label{figure6}
\end{figure*} 
\begin{table*}[t]
\renewcommand{\arraystretch}{1}
\centering
\caption{The number of patients in each cluster for DESLORATADINE drug}
\label{table6}
\centering
\begin{tabular}{|c|c|c|c|}
\hline
     & Cluster 1 (similar) & Cluster 2 (Non-similar) & Cluster 3 others  \\
\hline
 Euclidean Metric & 513 & 114	& 361 \\
\hline
Minkowski, $p$=3	& 578 & 89	& 321 \\
\hline
Manhattan Metric	& 602 & 89	& 304 \\
\hline
Hamming Metric	& 579 & 75	& 334 \\
\hline
PQ-Gram Metric	& 409 & 164	& 415 \\
\hline
Edit Distance Metric	& 284 & 332	& 372 \\
\hline
\end{tabular}
\end{table*}
%\clearpage
\begin{table*}[t]
\renewcommand{\arraystretch}{1}
\centering
\caption{The number of patients in each cluster for DOXEPIN drug}
\label{table7}
\centering
\begin{tabular}{|c|c|c|c|}
\hline
     & Cluster 1 (similar) & Cluster 2 (Non-similar) & Cluster 3 others  \\
\hline
 Euclidean Metric & 23 & 3	& 16 \\
\hline
Minkowski, $p$=3	& 81 & 7	& 17 \\
\hline
Manhattan Metric	& 23 & 3	& 16 \\
\hline
Hamming Metric	& 23 & 3	& 16 \\
\hline
PQ-Gram Metric	& 15 & 15	& 12 \\
\hline
Edit Distance Metric	& 16 & 15	& 11 \\
\hline
\end{tabular}
\end{table*}

\subsubsection{Clustering the Distances}
The results in Table \ref{table3} and Table \ref{table5} show the similarities and closest distances between patients using the previously mentioned metrics. The following step of this work has been to use a  clustering method to verify our results, to give the first insight on how the data looks like and to find which distance metric can represent similar distances better than the others. The clustering process has been also used to show whether all the similar distances in Tables \ref{table3} and \ref{table5} fall in one cluster or are distributed over all or some clusters. In this work, we used the $k$-means method and we chose the number of clusters to be equal to three clusters. For the first drug, two figures are reported to show the clusters of patients (Fig.~\ref{figure6a} and Fig.~\ref{figure7b}) using Euclidean and $pq$-gram distance metrics (a metric from each group of metrics). Fig.~\ref{figure8c} and Fig.~\ref{figure9d} show the clusters of patients using the same metrics for the second drug.\\
Since the $k$-means algorithm is known to be biased by the starting positions, it needs to be re-run more than once. As a result, we may get more than one outcome. The figures of the clusters represented in this work are those resulting from the most frequent clustering (the majority vote, in our experiments 10 times running). In order to distinguish between the clusters, we report Table \ref{table6} and Table \ref{table7} that contain the number of patients in each cluster for the first and second drug, respectively.
\emph{Cluster}1 in the tables contains the number of all the patients who are similar to each other, whereas \emph{cluster}2 contains all the patients who have large distances between each other. The remaining patients are grouped in \emph{cluster}3.
 
\section{DISCUSSION}
\label{sec:section4}
In this paper, six different distance metrics are applied to the THIN database for DESLORATADINE and DOXEPIN drugs. Our main objective is to find which metrics are useful for measuring distances in THIN data, with an emphasis on the $pq$-gram metric which is designed for hierarchical data like the read codes in THIN. We have implemented the distance metrics using two different types of data structures and compared their results. The two data structures are the tree-like structure of the group of $pq$-gram and Edit Distance metrics as shown in Fig.~\ref{figure4} and the frequency table or matrix for the group of geometric and Hamming metrics as shown in Table~\ref{table2}. The distance metrics have been applied to the data and generally, the results revealed that these metrics produced good similarity distances between patients' data. Regarding the $pq$-gram, the distances depend mainly on the number of intersected $pq$-grams between two trees as well as the values of the parameters $p$ and $q$. Choosing the correct values of $p$ and $q$ is a matter of tradeoffs. In \cite{srivastava2010}, Srivastava et al. analysed the sensitivity of $pq$-gram distances with the values of $p$ and $q$ and concluded that increasing $p$ relative to $q$ implies that more importance is being given to the ancestors than to the children of the trees, i.e. two nodes are considered to be the same only when they share $p$ common ancestors.\\
Thus, in our case the smaller the value of $p$ relative to $q$, the more probability of finding the intersected $pq$-grams between two trees and the more importance is given to the data rather than the structure of the trees. Based on that, the results in the seventh column in Table \ref{table3} and Table~\vref{table5} are better compared to the results of the eighth column of the same tables. In general, the $pq$-gram metric is not the best metric compared to the other metrics as it depends on many parameters ($p$, $q$ and the tree structure), but it could highlight some similar patients and measure the similarity between their data as shown in Table \ref{table3} (e.g. patients a670605Up, a670602uS and  patients a67340327, a681001KN). On the other hand, Table \ref{table5} contains some non-similar distances produced by the $pq$-gram and Edit Distance metrics, for example the two patients (g989501KB and a999104cU) have the normalised distance equal to 1 which means there is no similarity between both patients' data. The reason behind that could be the lack of data for the DOXEPIN drug. That is to say, the more data available the more probability of having similar data for patients in the THIN database.\\ 
 After finding all the distances using the chosen metrics, we verified our results by considering all the population of patients for each drug and by checking weather these distance metrics discriminate sufficiently using clustering the patient population. Fig.~\ref{figure6a}, Fig.~\ref{figure7b}, Fig.~\ref{figure8c} and Fig.~\ref{figure9d} show the results of clustering using the $k$-means algorithm. The latter is the simplest clustering method and requires the number of clusters to be known in advance. In this work, we chose the number of clusters to be equal to 3. However, more proper data analysis is required for future work and more than three clusters might be considered.
 The clusters have been plotted using the \emph{clusplot} function from R software which is representing all the observations by points in the plots using the principal component analysis \cite{Maechler2012}. PCA is used in the data set for the purpose of visualisation and no feature selection has been carried out. The clusters are labeled using numbers (1, 2 and 3) as shown in Fig.~\ref{figure6} and the geometric and Hamming metrics discriminate successfully on the population for both drugs. We chose only two figures for each drug, one for each group of metrics. Table \ref{table6} and Table~\vref{table7} show the number of patients in each cluster. The patients in Table \ref{table3} are grouped in \emph{cluster}1 for all the metrics used, while the patients in Table \ref{table5} are grouped in \emph{cluster}1 for the geometric and Hamming distance metrics only. In contrast, the distances for the same patients using $pq$-gram and Edit Distance metrics have a very poor similarity. Thus \emph{cluster}1 for the both metrics contains some similar distances other than those in Table \ref{table5}. The reason behind that probably is the lack of data related to the second drug.
In general, \emph{cluster}1 in Table \ref{table6} and Table \ref{table7} contains the similar patients who have all or some medical events related to the drug in common, while \emph{cluster}2 contains the non-similar ones. All the other patients who are not in \emph{cluster}1 or \emph{cluster}2 are grouped in \emph{cluster}3 as shown in Fig.~\ref{figure6a}, Fig.~\ref{figure7b}, Fig.~\ref{figure8c} and Fig.~\ref{figure9d}.

\section{SUMMARY AND CONCLUSION }
\label{sec:section5}
Two groups of distance metrics have been considered for two kinds of data structures from the THIN longitudinal health-care database, and then compared. The comparison is done by firstly looking at whether each metric can measure any distances and if all the metrics find the same similar patients with the same distances, and secondly by clustering the whole population of patients to find if the metrics sufficiently discriminate those patients. The results show that the two groups of metrics worked successfully in finding similar distances for similar patients and group all them in one cluster when clustering using the $k$-means algorithm.\\
In conclusion, the $pq$-gram metric might not be the best metric for THIN data, but it can measure similar distances and group them in one cluster. That is to say, it highlighted some known medical events related to the drugs been taken, for example the cough and itchy eye symptoms related to DESLORATADINE drug. As each group of metrics depends on different data structures and in order to choose the appropriate distance measure for the THIN data, we may need an appropriate structure of the data: for example, a mixed data structure from both the hierarchical and non-hierarchical data. By making the tree structure for all the levels of read codes, the distances can be calculated for read codes only. As a result of that, the $pq$-gram could find the related medical codes to each other in a better way.

\bibliographystyle{IEEEtran} 
\bibliography{References}
\end{document}